\documentclass{emulateapj}

\usepackage{apjfonts}
\usepackage{natbib}

\newcommand{\del}[2]%
{\frac{\mathrm{d}{#2}}{\mathrm{d}{#1}}}
\newcommand{\Del}[2]%
{\frac{\mathrm{D}{#2}}{\mathrm{D}{#1}}}
\newcommand{\ddel}[2]%
{\frac{\mathrm{d}^2{#2}}{\mathrm{d}{#1}^2}}

\newcommand{\pdel}[2]%
{\frac{\partial{#2}}{\partial{#1}}}
\newcommand{\pddel}[2]%
{\frac{\partial^2{#2}}{\partial{#1}^2}}

\newcommand{\simgt}{\lower.5ex\hbox{$\; \buildrel > \over \sim \;$}}
\newcommand{\simlt}{\lower.5ex\hbox{$\; \buildrel < \over \sim \;$}}

\def\Rnum#1{\uppercase\expandafter{\romannumeral #1}}
\def\rnum#1{\expandafter{\romannumeral #1}}

\slugcomment{Draft version \today}

\shorttitle{Gravitational Wave Background from Pop III Stars}
\shortauthors{Suwa et al.}

\begin{document}

\title{Gravitational Wave Background from Population III Stars}

\author{
Yudai Suwa\altaffilmark{1},
Tomoya Takiwaki\altaffilmark{1},
Kei Kotake\altaffilmark{2,3},
and Katsuhiko Sato\altaffilmark{1,4}
}
\altaffiltext{1}{
Department of Physics,School of Science, the University of Tokyo, 7-3-1 Hongo,
Bunkyo-ku, Tokyo 113-0033, Japan}
\altaffiltext{2}{
Division of Theoretical Astronomy, National Astronomical Observatory of Japan,
Mitaka, Tokyo 181-8588, Japan}
\altaffiltext{3}{Max-Planck-Institut f\"{u}r Astrophysik, Karl-Schwarzshild-Str.
 1, D-85741, Garching, Germany}
\altaffiltext{4}{Research Center for the Early Universe,
School of Science, the University of Tokyo,7-3-1 Hongo,
Bunkyo-ku, Tokyo 113-0033, Japan}
\email{suwa@utap.phys.s.u-tokyo.ac.jp}

\begin{abstract}
We estimate the gravitational wave (GW) background from Population III (Pop III) stars
using the results from our hydrodynamic simulations (Suwa et al. 2007). 
We calculate gravitational waveforms from
matter motions and anisotropic neutrino emissions for single Pop III
stars.
We find that the GW amplitudes from matter motions are dominant until 
just after bounce,
but those from neutrinos dominate later
on at frequencies below $\sim 10$ Hz in the GW spectrum.
Computing the overall signal produced by the ensemble of such
Pop III stars, we find that the resultant density parameter of the GW
background peaks at the amplitude of $\Omega_\mathrm{GW}\sim 10^{-10}$ in the
frequency interval $\sim 1-10$ Hz. We show that such signals,
depending on the formation rate of Pop III stars,
can be within the detection limits
of future planned interferometers such as
DECIGO and BBO in the frequency interval of $\sim 0.1-1$ Hz.
Our results suggest that the detection of the GW background from Pop III stars can
be an important tool to supply
the information about the star formation history in the early universe.
\end{abstract}


\section{Introduction}

The observation of gravitational waves (GWs) is one of the most important tasks
for exploring the less well known parts of the Universe.
Already several ground-based laser interferometers (TAMA300, LIGO, and GEO600) are
now operating and taking data
in the frequency range of 10 Hz - 10 kHz,
where rapidly-collapsing stellar objects accompanied with strong gravity
 such as neutron stars (NSs) and black holes (BHs),
are one of the most promising sources of GWs
\cite[see, e.g.,][for reviews]{new03,kotarev}.
The Laser Interferometer Space Anntena (LISA), covering $10^{-4}-10^{-2}$ Hz,
 will be launched in the near future,
and, moreover, future space missions such as DECIGO \citep{seto01} and BBO \citep{unga05},
targeting the $\sim$ 0.1 Hz regime, are being planned.
Possible targets of these experiments in the low frequency region are gravitational
 wave backgrounds (GWBs) from both astrophysical and cosmological origins.
In particular the low-frequency window is thought to be indispensable
for the detection of the primordial GWB generated during the inflationary epoch
\citep{magg00}.

In addition, various possible astrophysical sources of GWBs have been investigated.
Among them are core-collapse supernovae \citep{arau04},
inspiral and coalescence of compact binaries
\citep{farm03}, rotating NSs with non-axisymmetric deformations \citep{ferr99b}, gamma-ray bursts \citep{hira05},
and NSs with phase transitions
\citep{sigl06} (see references therein).
Recently, \citet{buon05} pointed out that the contribution from Population III
(hereafter Pop III) stars is most important because it could be
 as large as the inflationary GWB around
0.1 Hz where DECIGO and BBO are most sensitive
\cite[see also][]{sand06}.

Here, Pop III stars are the first stars, which formed out
of a pristine, metal free gas produced by the big bang \cite[for reviews, see, e.g.][]
{brom04, glov05}.
Recent simulations indicate that the initial mass function (IMF) of Pop III stars is expected
to have been dominated by very massive stars with mass $\simgt 100M_\odot$
\citep{naka01,abel02,brom02b,omuk03}.
These very massive stars encounter the electron-positron pair creation
instability after central carbon burning, which reduces
the thermal energy and induces gravitational collapse.
For stellar mass of less than $\sim 260 M_\odot$, rapid nuclear burning releases large
energy sufficient to entirely disrupt the star as pair-instability supernova.
More massive stars, which also encounter pair-instability, are so tightly bound that the
fusion of oxygen is unable to reverse infall.
Such stars are thought to collapse to BHs \citep{bond84,fryer01},
which we pay attention to in this letter.
So far, there have been only a few hydrodynamic simulations studying
the gravitational collapse of the BH forming Pop III stars
\citep{fryer01, naka06, suwa07a,liu07}.

Due to the lack of Pop III star collapse simulations,  \citet{buon05}
had to calculate the GWB with an assumption that the GW spectrum
of a single Pop III star has the same shape as the one from an ordinary
core-collapse supernova. They employed the result of \citet{muel04} as the template.
In order to determine the normalization of the GW amplitudes,
they simply amplified the GW amplitudes of the supernova by a factor
of $\sim 1000$, relying on the results by \citet{fryer01}.
Needless to say, these assumptions and treatments can be validated by
a consistent treatment, namely by hydrodynamic collapse simulations of Pop
III star with the analysis of the GW waveforms.
As for the mass and formation history of Pop III stars, \citet{buon05}
assumed that
all stars formed with the same mass of $300 M_\odot$ and at the same redshift as
$z \sim 15$.
This assumption is improved by \citet{sand06}, who calculated GWB from Pop III stars
with the cosmic star formation history in the framework of hierarchical structure
formation.
However, \citet{sand06} also employed the same spectrum of a single Pop III star
as \citet{buon05}.

The purpose of this letter is to give an estimate of the GWB spectrum from
Pop III stars by calculating
the GW waveforms based on the results of hydrodynamical
core-collapse simulations of Pop III stars.
This letter is organized as follows:
In the next section the numerical model is briefly described.
In \S3, we calculate the gravitational wave signal and its spectrum of a single
Pop III star collapse.
In \S4, we present the numerical result of the gravitational wave background from
Pop III stars.
\S 5 is devoted to summary and discussion.

\section{Method}
The numerical methods are basically the same as 
 the ones in our previous paper \citep{suwa07a}.
With the ZEUS-2D code \citep{ston92} as a base for the hydro
 solver, we employ a realistic equation of state based on
relativistic mean field theory \citep{shen98} (see \citealt{kota03} for
 the implementation) and treat the neutrino 
cooling by a multiflavour leakage scheme, in which 6 species of neutrinos
 with pair, photo, and plasma processes by \cite{itoh89} in addition
to the standard charged current neutrino cooling reactions
 are included.
Spherical coordinates $(r,\theta)$ with logarithmic zoning in
the radial direction and regular zoning in $\theta$ are used.
One quadrant of the meridian section is covered with 300 ($r$)$\times$ 30
($\theta$) mesh points.
In our 2D calculations, axial symmetry and reflection symmetry across the
equatorial plane are assumed.
We also calculated some models with 60 angular mesh points,
but found no significant differences.
Therefore, we will report in the following the results obtained from the models
with 30 angular mesh points.

The initial condition is provided in the same manner of
\citet{suwa07a}. We produced hydrostatic cores of 300, 500, 700, and 900 $M_\odot$,
with the assumption of isentropic, whose values of entropy are taken
from \citet{bond84}, and with the constant electron fraction of $Y_\mathrm{e}=0.5$.
The supposed rotation law is cylindrical rotation, with the strength
of rotational energy taken to be 0.5\% of the gravitational energy in all models.

In this study, we estimate the gravitational wave emission from aspherical
mass motions via the Newtonian quadrupole formula of \citet{moen91}.
In addition, we compute the gravitational wave
strain from anisotropic neutrino emission employing the
formalism introduced by \citet{epst78}
and developed by \citet{muel97} and \citet{kota07}.
The GW emission from neutrinos is given as
\begin{equation}
\label{amplitude}
  D h^{TT}(t)=\frac{2G}{c^4}\int^t_{-\infty}\alpha(t')L_\nu(t')dt',
\end{equation}
where $D$ is the distance to the source, $h^{TT}$ is the transverse-traceless
and dimensionless metric strain, $G$ is the gravitational constant, $c$ is the
speed of light, $\alpha(t)$ is the time-dependent
neutrino-anisotropy parameter, and $L_\nu(t)$ is the total neutrino luminosity.
For the estimation of $\alpha$, neutrinos are assumed to be emitted radially in
each angular bin. 

\section{Gravitational Wave of a single Pop III star collapse}
\begin{figure}[tbp]
\plotone{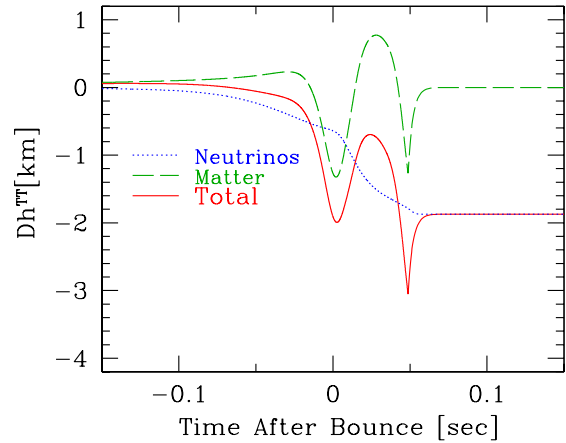}
\caption{The gravitational wave strain, $h^{TT}$, times the distance to the Pop III star, $D$,
versus time after bounce (in seconds). 
Shown are the neutrino (blue dotted) and matter motion (green dashed) components along
with their sum (red solid).
}
\label{time_ev}
\end{figure}

In this section, we discuss the GW emission from a single 300$M_\odot$
Pop III star collapse.
%
Figure \ref{time_ev} depicts the strain versus time after bounce for a $300M_\odot$ star.
The matter contribution dominates $h^{TT}$ during first 10 ms after bounce.
Afterward, the neutrino part begins to contribute because
the thermal shock occurs and the neutrino luminosity increases
in the hot region behind the shock wave.
At about 70 msec after bounce, the neutrino contribution
converges to a constant value and the matter contribution goes to zero.
This represents the epoch of the BH formation. 
After BH formation, we do not calculate the GW emission from matter but dump
with the timescale of
light crossing time $\sim O(R_\mathrm{BH}/c)$, where $R_\mathrm{BH}$ is the radius of
the BH.
This procedure does not affect the discussion of the following section
because the matter contribution of total GW is only in the high frequency region,
which is not the main point of this letter.
The total energy emitted in gravitational waves is $\sim
2\times10^{50}$ erg, which is smaller than the result of \citet{fryer01}
by a factor of 10.
The discrepancy seems to come from the difference of initial angular momentum distribution.

\begin{figure}[tbp]
\plotone{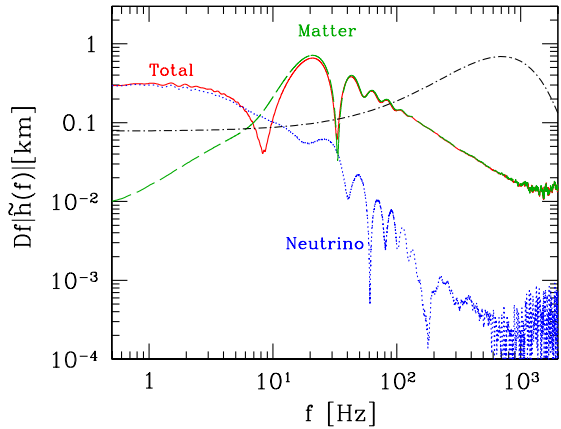}
\caption{
Spectral distributions of the gravitational waves from anisotropic neutrino emissions
(blue dotted line), matter motions (green dashed line), and total (red solid line).
Due to phase cancellations between the matter and neutrino components, the amplitude of
the combined spectrum is smaller than the dominant components in certain frequency regions.
The dominance of the neutrino-oriented gravitational wave component at frequencies below 
$\sim 10$ Hz is clearly seen.
The dot-dashed line represents the spectrum employed in previous works \citep{buon05,sand06}.
}
\label{fourier}
\end{figure}

Now we move on to discuss the features of the waveforms by the spectral analysis.
From Figure \ref{fourier}, one can see the dominance of the neutrino-oriented gravitational
waves at frequencies below $\sim$ 10 Hz.
This is because the waveform from neutrinos shows the slower time-variability in comparison with the one from matter motion
(see Figure \ref{time_ev}).
On the other hand, the contribution of matter motion dominates the
spectrum in the higher frequency region.
The peak amplitudes from matter motion and the
frequencies of 10s of Hz are roughly at the same characteristic
amplitude and frequency produced by the collapse of the star to form a BH, which are
evaluated in \citet{thor87}, with the remnant mass of $\sim 100M_\odot$.

For comparison, in Figure \ref{fourier} we also plot the spectrum assumed in previous works
(dot-dashed; \citealt{buon05,sand06}).
The result of our calculation is very different from their spectrum.
This is due to consideration of hydrodynamical features.
Pop III stars have higher temperature prior to collapse, which leads
core-bounce by the (gradient of the) thermal pressure and not by the nuclear forces
as in ordinary supernovae \citep{fryer01,suwa07a}.
Such thermal bounce makes the central density lower at the time of bounce
($O(10^{12})$g~cm$^{-3}$) so that
the dynamical timescale
($\propto \rho^{-1/2}$)
becomes longer, leading to the smaller time-variability in the
waveforms and the smaller typical frequencies (a few 10 Hz) where
the matter contribution peaks, which are in the 1 kHz regime for ordinary supernovae.

It should be noted that current numerical simulations 
only encompass a few seconds at most and 
do not cover the strain spectrum below a fraction of a Hertz.
For such a low frequency region, we apply the zero-frequency
limit \citep{smar77,epst78} to extract the GW waveforms as done in \cite{buon05}.

Finally, we discuss the GW emission from stars with different mass models.
It is noted that the qualitative features are common to all models.
In Table \ref{tab:gw}, we show the characteristic values of GW spectra.
It can be seen that the peak amplitude, $\tilde h_\mathrm{max}$, increases 
with the mass of the star, while
the peak frequency, $f_\mathrm{max}$, remains roughly the same.
The frequency where the neutrino and matter contributions become equal,$f_\mathrm{eq}$, 
decreases with increasing mass.
The GW amplitudes for the zero frequency limit,
$Df|\tilde h|(f\rightarrow0)$,
increases with the mass since the
luminosity of neutrinos becomes larger.

\begin{table}[htbp]
\begin{center}
\caption{Characteristic Quantities of Gravitational Wave}
\label{tab:gw}
\begin{tabular}{ccccc}
\hline
\hline
Mass~~  & $D\tilde h_\mathrm{max}$~~ & $f_\mathrm{max}$~~ &
$f_\mathrm{eq}$~~ & $Df|\tilde h|(f\rightarrow0)$ \\
($M_\odot$) & (km) & (Hz) & (Hz) & (km)\\
\tableline
300 & 0.66 & 21 & 7.8 & 0.30 \\
500 & 0.88 & 19 & 6.0 & 0.46 \\
700 & 0.86 & 23 & 4.4 & 0.38 \\
900 & 0.99 & 18 & 4.0 & 0.54 \\
\tableline
\end{tabular}
\end{center}
\end{table}

\section{Gravitational Wave Background from Pop III Stars}

We are now in a position to discuss the contribution of GWs from Pop III
stars to the background radiation.
According to \cite{phin01}, the sum of the energy densities radiated by a large
number of independent Pop III stars at each redshift is given by the density
parameter $\Omega_\mathrm{GW}(f)\equiv\rho_c^{-1}(d\rho_\mathrm{GW}/d\log f)$ as
\begin{equation}
  \label{eq:omega}
  \Omega_\mathrm{GW}(f)=\frac{16\pi^2c}{15G\rho_c}\int\frac{dz}{1+z}
  \left|\frac{dt}{dz}\right|\psi(z)\int dm\ \phi(m)D^2f'^3|\tilde h(f')|^2,
\end{equation}
where $\rho_c$ is the critical density ($3H_0^2/(8\pi G)$), 
$\psi(z)$ is the star formation rate (SFR),
$\phi(m)$ is the initial mass function (IMF) of Pop III stars, 
and $f'$ is the red shifted frequency, $(1+z)f$.
We employ model 2b of \citet{sand06} for the SFR,
which is appropriate for very massive stars, from $270M_\odot$ to $500M_\odot$.
As for the IMF, we employ the same parameterization as \citet{sand06},
$\phi(m)\propto m^{-2.3}$,
which is normalized by $\int dm\phi(m)=1$.
The cosmological model enters with $|dt/dz|=[(1+z)H(z)]^{-1}$ and, for a flat geometry,
$H(z)=H_0[\Omega_\Lambda+\Omega_m(1+z)^3]^{1/2}$.
We use the parameters $\Omega_\Lambda=0.73, \Omega_m=0.27$, and $H_0=100\ h_0
$km s$^{-1}$ Mpc$^{-1}$ with $h_0=0.71$ \citep{sper07}.

\begin{figure}[t]
\plotone{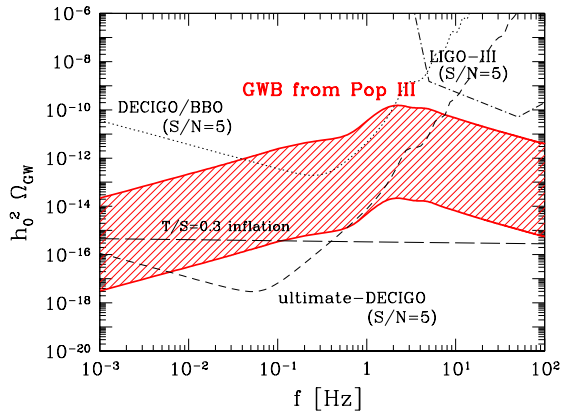}
\caption{
The energy density parameter of gravitational wave background (GWB).
The upper line of the shaded region is the result of our calculation.
The lower line is for the renormalized SFR of $f_{III}\sim10^{-5}$.
In low frequency region ($f \simlt $0.1Hz), $\Omega_\mathrm{GW}\propto f$ because 
$f |\tilde h|$ is constant (see Figure 2) and
the integrand of Eq. (2) is $f^3 |\tilde h|^2 \propto f$ in this region.
The horizontal dashed line shows the GWB produced during slow-roll inflation assuming
$T/S=0.3$ for the ratio of the tensorial and scalar contributions to the cosmic 
microwave background radiation anisotropy and no running of the tensorial 
power-law index, which is evaluated from Eq. (6) of
\citet{turn97}.
The sensitivity curves of space- and (correlated) ground-based detectors 
are taken from \citet{kudo06} and \citet{buon03}.
}
\label{omega}
\end{figure}

In Figure \ref{omega}, the calculated $\Omega_\mathrm{GW}$ is plotted with the sensitivity
curves of future detectors.
The upper edge of the red shaded region is obtained by our calculation
with model 2b of \citet{sand06}.
This corresponds to almost the
upper limit of the SFR because the stars with the mass range considered
in this model entirely collapse to BHs and do not contribute to the
chemical evolution of their environment.
In this way, a high SFR is obtained without metal overproduction.
Meanwhile, the lower limit is the same shifted downwards
by a factor 7000, which corresponds to the baryon fraction of Pop III stars of $10^{-5}$,
suggested by Pop III star formation theory (Omukai, private communication).
It can be seen that the contributions from neutrinos,
which dominates in low frequency region ($\sim 1$ Hz), 
are within the detection limit of the planned
detectors DECIGO and BBO.
In contrast, detection by LIGO III, which is a ground based detector, 
may prove difficult
because the amplitude in the high frequency region is smaller than the
current estimated detection limit.
Even if the SFR of Pop III stars is as small as in the case of a 
baryon fraction of $\sim 10^{-5}$,
the GWB from Pop III stars might be within the detection limit of ultimate-DECIGO.
The GWB spectrum predicted in this letter is larger than previous works 
by a factor of 40
in the low frequency region
(see Figure 8 of \citealt{sand06}, 
which represents larger amplitude than \citealt{buon05}).
This is due to the difference of a GW spectrum of a single Pop III star collapse
as already mentioned in \S3.
Depending on the baryon fraction of Pop III stars,
the GWB from Pop III stars might give a strong contribution, masking the
GWB generated in the inflationary epoch 
(the horizontal dashed line).
We furthermore point out that 
the amplitude of the GWB is highly dependent on the SFR but less sensitive
to the index of IMF because the GW emission of different mass Pop III stars is similar
in the mass range focused on in this letter.

\section{Summary and Discussion}
In this letter, we have presented the GWB spectrum from Pop III stars
based on the hydrodynamic core-collapse simulations of Pop III stars
\citep{suwa07a}. Extracting the GWs contributed from mass motions
and anisotropic neutrino radiation,
we have found that the GW emission from neutrinos is dominant over the ones from
 matter at low frequencies. Using the spectrum of single Pop III
 stars, we have calculated the GWB
spectrum by summing up the contribution from individual stars.
We have shown that the amplitudes from Pop III stars
might be large enough to be  detected by interferometers such as DECIGO
and BBO.

The predicted GWB is in the sensitivity range of future planned
detectors so that the Pop III stars might limit the sensitivity
of such detectors.
This is analogous to the discussion about the the Galactic population of
close white dwarf binaries, which could produce a confusion-limited
 GW signal at the lower end of the sensitivity band of LISA \citep{hils90}.

It is noted that the Newtonian simulation in the paper
\citep{suwa07a} is nothing
but an idealized study to describe the dynamics with BH
formation. It is by no
 means definitive. To explore these phenomena in more detail will
 require fully general relativistic simulations and much more better
 neutrino transfer.
However, in comparison with the forgoing GWB studies
simply using the GW waveform of ordinary core-collapse supernovae
as a template,
we have found significant differences in the properties of the
waveforms, which made it possible to realize the potential importance
of Pop III stars as a contribution to the GWB.
As mentioned, our results show
enhancements in the GWB amplitudes from
0.1 to 1 Hz compared to earlier works, 
where the currently planned detectors are most sensitive.

Finally, we shall briefly discuss the uncertainties of our models.
Since little is known about the angular momentum distributions of Pop III stars,
we took the initial rotation rate of extremely rapidly rotating cores of 
massive stars \citet{hege00}
as a reference suggested by the study that Pop III stars could rotate
rapidly due to the
insufficient mass-loss in the main sequence stage 
\citep{hege03a}.
Obviously more systematic
studies changing the initial rotation rates and profiles are required.
Next, the IMF of Pop III stars is also uncertain.
It is true that the mass range employed here ($300 - 1000 M_\odot $) is preferred by
recent studies of Pop III star formation, but that the possible
modification of the IMF leads to large changes of the predicated GWB spectrum
\citep[see, e.g,][]{schn00,arau02}.
This means that the GWB could be a powerful tool to investigate the 
IMF. 
Although the baryon fraction of Pop III stars taken here is indeed not too 
large for the explanation of the infrared 
background excess by the UV photons from Pop III stars 
\citep{sant02,dwek05}, there still remains large uncertainty.
The last uncertainty is the redshift where the SFR
 becomes maximum. It should be mentioned that the frequencies where
 the GWB from Pop III stars peaks are sharply dependent on this
 redshift. Better correlation analysis between multiple 
spaced-based interferometers should be very helpful, because this can improve 
the sensitivity of GWB detection \citep{kudo06}.
All these results suggest that 
 detections of GW background from Pop III stars can be an important
tool to supply information about the 
formation history of Pop III stars.

\acknowledgments
Y.S. would like to thank T. Hiramatsu, S. Kinoshita,
K. Omukai, E. Reese, S. Saito, M. Shibata and A. Taruya 
for helpful discussion.
Numerical computations were in part carried on VPP5000
and general common use computer system at the center for
Computational Astrophysics, CfCA, the National
Astronomical Observatory of Japan.
This study was supported in part by the Japan Society for
Promotion of Science (JSPS) Research Fellowships,
Grants-in-Aid for the Scientific Research from the Ministry of Education,
Science and Culture of Japan (No.S19104006).

\end{document}